\documentclass{emulateapj}

\shorttitle{Accretion-powered Stellar Winds}
\shortauthors{Matt \& Pudritz}

\begin{document}

\title{Accretion-powered Stellar Winds \\ as a Solution to the Stellar Angular Momentum Problem}

\author{Sean Matt\altaffilmark{1,2,3} and Ralph E. Pudritz\altaffilmark{1}}

\altaffiltext{1}{Physics and Astronomy Department, McMaster University,
Hamilton, ON L8S 4M1, Canada; seanmatt@virginia.edu,
pudritz@physics.mcmaster.ca.}

\altaffiltext{2}{Current address: Department of Astronomy, University of
Virginia, P.O. Box 3818, Charlottesville, VA 22903.}

\altaffiltext{3}{Levinson/VITA Fellow, University of Virginia.}

\begin{abstract}

We compare the angular momentum extracted by a wind from a
pre-main-sequence star to the torques arising from the interaction
between the star and its Keplerian accretion disk.  We find that the
wind alone can counteract the spin-up torque from mass accretion,
solving the mystery of why accreting pre-main-sequence stars are
observed to spin at less than 10\% of break-up speed, provided that
the mass outflow rate in the stellar winds is $\sim 10\%$ of the
accretion rate.  We suggest that such massive winds will be driven by
some fraction $\epsilon$ of the accretion power.  For observationally
constrained typical parameters of classical T-Tauri stars, $\epsilon$
needs to be between a few and a few tens of percent.  In this
scenario, efficient braking of the star will terminate simultaneously
with accretion, as is usually assumed to explain the rotation
velocities of stars in young clusters.




\end{abstract}


\section{Introduction} \label{sec_intro}

Pre-main-sequence stars surrounded by Keplerian disks accrete
substantial amounts of angular momentum along with infalling matter
and energy.  Classical T-Tauri stars (CTTSs) are widely understood to
be low-mass ($\la 2 M_\odot$) pre-main-sequence stars with ages
ranging from a few times $10^{5}$ to a few million years and represent
the latest stages of protostellar accretion.  The typical accretion
torque on these stars is sufficient to spin them up to break-up speed
in much less than $10^6$ yrs \citep{hartmannstauffer89}.  The fact
that many CTTSs (the ``slow rotators'') spin at $\la 10$\% of break-up
\citep[][]{bouvierea93} and have ages longer than their spin-up times,
suggests that they are in spin equilibrium, wherein they somehow rid
themselves of accreted angular momentum and thereby maintain a net
zero torque.  Furthermore, in order to explain the distribution of
rotational velocities of stars in young clusters, it is generally
believed \citep[][for a review]{edwardsea93, bodenheimer95} that
rotational braking of the star becomes inefficient when accretion
ceases.

The leading explanation for angular momentum loss during accretion,
referred to as ``disk locking'' \citep{ghoshlamb78, konigl91,
shuea94}, requires a significant spin-down torque on the star arising
from a magnetic connection between the star and disk.  However,
\citet[][and references therein]{mattpudritz05} discussed several
severe problems with the disk locking scenario, most notably that the
stellar magnetic field topology should be largely open, rather than
connected to the disk.  The presence of open stellar field lines
allows for, and may be caused by, a stellar wind, and the immediate
question is whether a wind along these open lines carries away enough
angular momentum to counteract the accretion torque
\citep{hartmannstauffer89, toutpringle92}.


In this Letter, we report (in \S \ref{sec_equilib}) that protostellar
winds can remove accreted angular momentum, even for slow stellar
rotation rates, provided that the stars have large mass loss rates.  We
propose that the energy driving the stellar wind derives from
accretion power (\S \ref{sec_power}), and this also explains the
apparent connection between efficient braking and accretion.  In
section \ref{sec_synthesis}, we combine the results of many studies of
the star-disk interaction to give a complete picture for the flow of
matter and angular momentum near the star.

\section{Spin Equilibrium} \label{sec_equilib}

We consider long-term torques, averaged over $\sim 10^4$ yr (i.e.,
much less than spin-up/down times).  The approximation of a
steady-state and the adoption of global, axisymmetric magnetic fields
are thus acceptable, even though the magnetic structure, winds, and
accretion properties are variable and probably not axisymmetric, on
much shorter timescales.  The torque on the star, due to the accretion
of disk matter, is \citep[e.g.,][]{mattpudritz05}\footnote{We have
neglected a term proportional to the spin rate of the star, but eq.\
(\ref{eqn_tacc}) is valid for spin rates well below breakup speed.}
\begin{eqnarray}
\label{eqn_tacc}
\tau_{\rm a} = \dot M_{\rm a} \sqrt{G M_* R_{\rm t}},
\end{eqnarray}
where $\dot M_{\rm a}$ is the mass accretion rate, $G$ is the
gravitational constant, $M_*$ is the mass of the star, and $R_{\rm t}$
is the location of the inner edge of the disk, from which material
essentially free-falls onto the stellar surface \citep{konigl91}.  In
the following, we compare this accretion torque with the torque
originating from a stellar wind.

X-ray observations \citep{feigelsonmontmerle99} and magnetic field
measurements \citep[][]{johnskrull3ea99, smirnovea03} of CTTSs reveal
the presence of hot coronae and dynamically important fields.
Together with the rotation rates, these observations suggest that
CTTSs drive stellar winds by coronal thermal pressure (similar to the
Sun) and that magneto-centrifugal effects may also play a role.
Furthermore, \citet{dupreeea05} recently reported evidence for hot
($\sim 3 \times 10^5 $K), fast ($\sim 400$ km s$^{-1}$) stellar winds
from two CTTSs.  Thus, we believe it is appropriate to adopt standard
magnetohydrodynamic (MHD) wind theory \citep[e.g.,][]{weberdavis67,
mestel68, mestel84, sakurai85, kawaler88} for these systems.  In this
case, the torque on the star, due to angular momentum lost to the
wind, is given by \citep[e.g.,][]{mestel84}\footnote{Equation
(\ref{eqn_twind}) is valid for any magnetic geometry.
\citet{kawaler88} used a different formulation for a dipolar magnetic
field, but this was a misinterpretation of eq.\ (12) in Mestel (1984),
which requires stellar surface values of density and velocity, instead
of the total mass outflow rate $\dot M_{\rm w}$.}.
\begin{eqnarray}
\label{eqn_twind}
\tau_{\rm w} = -\kappa \dot M_{\rm w} \Omega_* R_*^2 (r_{\rm A} / R_*)^2,
\end{eqnarray}
where $\dot M_{\rm w}$ is the mass loss rate in the stellar wind,
$\Omega_*$ is the angular spin rate of the star, $R_*$ is the stellar
radius, and $r_{\rm A}$ is the Alfv\'en radius, defined as the
location where the wind velocity reaches the local Alfv\'en speed.
The dimensionless factor of order unity $\kappa$ takes into account
the geometry of the wind ($\kappa = 2/3$, for a spherically symmetric
wind).  In essence, wind theory tells us that magnetized stars spin
``with their arms out,'' and the resulting spin-down torque depends
most strongly on the length of their lever arm, $r_{\rm A}$.

Assuming that a spin-down torque arising from a disk connection and
spin-up due to contraction are negligible, the equilibrium spin rate
of the star is determined by the balance of accreted angular momentum
with the spin-down torque from the stellar wind
\citep{hartmannstauffer89}.  By equating $\tau_{\rm a} = -\tau_{\rm
w}$ (eqs.\ [\ref{eqn_tacc}] and [\ref{eqn_twind}]), the equilibrium
stellar spin rate is
\begin{eqnarray}
\label{eqn_feq}
f_{\rm eq} \approx 0.09
     \left({{{\kappa} \over {2/3}}}\right)^{-1}
      \left({{{R_{\rm t} / R_*} \over {2}}}\right)^{1/2}
      \nonumber \\ \times
      \left({{{r_{\rm A} / R_*} \over {15}}}\right)^{-2}
      \left({{{\dot M_{\rm w}/\dot M_{\rm a}} \over {0.1}}}\right)^{-1},
\end{eqnarray}
where we have expressed the spin rate as a fraction of the break-up speed,
$f \equiv \Omega_* R_*^{3/2} (G M_*)^{-1/2}$.  It is immediately
evident that stellar winds alone are capable of keeping CTTSs spinning
well below the break-up rate, provided that they drive powerful (i.e.,
large $\dot M_{\rm w}$) winds and have a long magnetic lever arm.

This result depends most strongly on the length of the lever arm,
which is an uncertain parameter.  The usual analytic calculation of
$r_{\rm A} / R_*$ \citep[e.g.,][]{kawaler88, toutpringle92} is not
very reliable, as it employs a one-dimensional formulation (instead of
the two-dimensional problem here), and it depends strongly on the
assumed magnetic geometry and wind speeds.  However, the analytical
result is still useful because it tells us $r_{\rm A} / R_*$ depends
on the ratio $B_*^2 R_*^2 / \dot M_{\rm w}$ (where $B_*$ is the field
strength at the stellar surface), for a given magnetic geometry and
wind speeds.  Using the well-studied example of the solar wind, we can
get an initial estimate of $r_{\rm A}$, as follows.  First we assume
that CTTS wind speeds are within a factor of a few of solar wind
values, which we expect from the similar escape speeds and is
supported by observations.  Second, for a lack of information to the
contrary, we assume that the magnetic geometry in CTTS winds is also
not too different from solar.  Now, using the observational limit on
the large-scale (dipole) component of CTTS magnetic fields of $B_*
\sim 200$ G \citep{johnskrull3ea99, smirnovea03}, and assuming $R_* =
2 R_\odot$, the ratio $B_*^2 R_*^2 / \dot M_{\rm w}$ is equal to the
solar wind value when $\dot M_{\rm w} \sim 2 \times 10^{-9} M_\odot$
yr$^{-1}$.  Remarkably, this is approximately 10\% of typical observed
accretion rates \citep{johnskrullgafford02}.  Thus, for this value of
$\dot M_{\rm w}$, the lever arm length should be close to the solar
value of 12--16 \citep{li99} and larger if $\dot M_{\rm w}$ is
smaller.  Furthermore, this value of $r_{\rm A}$ is consistent with
numerical simulation results \citep[e.g.,][]{mattbalick04}, when
scaled for CTTS winds, even for rotation rates of 10\% of breakup.
Therefore, we believe that $r_{\rm A} / R_* \ga 15$ is reasonable, and
the fiducial value in equation (\ref{eqn_feq}) is justified.

The relationship $\dot M_{\rm w} \sim 0.1 \dot M_{\rm a}$ is
consistent with the stellar outflow rates reported by
\citet{dupreeea05}, as well as the large-scale mass outflow rates
inferred in these and younger systems \citep{koniglpudritz00}.  As
further support, a coronal wind with $\dot M_{\rm w} \la 10^{-9}
M_\odot$ yr$^{-1}$ is consistent with CTTS X-ray luminosities
\citep{decampli81}.  In the following section, we show that accretion
power is capable of driving massive stellar winds such as these.

\section{Accretion Power} \label{sec_power}

If the stellar wind alone counteracts the accretion torque, equation
(\ref{eqn_feq}) indicates that $\dot M_{\rm w}$ should be a substantial
fraction of $\dot M_{\rm a}$, which requires powerful wind driving.
The observations discussed in \S \ref{sec_equilib} indicate that CTTSs
have enhanced rotational, thermal, and magnetic energies in their
coronae, relative to the present day Sun, suggesting that CTTS winds
will be substantially more energetic and massive than the solar wind.
It is not yet clear, however, whether scaled-up solar-type activity
alone can drive high enough mass loss to satisfy equation
(\ref{eqn_feq}) \citep{decampli81, toutpringle92, kastnerea02}.
Instead, we propose that the stellar wind is powered by the energy
deposited on the star via accretion.  This scenario is supported by
observations of hot, stellar outflows \citep{beristain3ea01,
edwardsea03, ferrofontangomezdecastro03, dupreeea05}.

The details of the complicated interaction between the star and disk
are not important for tabulating the accretion power.  Instead, this
can be characterized as an inelastic process, wherein rotating disk
material attaches itself to the stellar magnetosphere at $R_{\rm t}$,
and eventually falls onto and becomes part of the star.  What matters
is the difference in the energy before and after this interaction.  In
particular, disk matter that falls from $R_{\rm t}$ to $R_*$ liberates
gravitational potential energy and also transfers its orbital kinetic
energy onto the star.  Some of this energy is added to the rotational
kinetic energy of the star (at a rate $\Omega_* \tau_{\rm a}$), and,
in spin equilibrium, is balanced by the work done by the stellar
rotation on the wind (at a rate $\Omega_* \tau_{\rm w}$).  The
remaining accretion power is\footnote{We have neglected terms
proportional to $f^2$, which are important only for fast rotation.}
\begin{eqnarray}
\label{eqn_la}
L_{\rm a} = 0.5 \dot M_{\rm a} v_{\rm esc}^2
     [1 - 0.5 R_*/R_{\rm t} - f (R_{\rm t}/R_*)^{1/2}],
%
%
\end{eqnarray}
where $v_{\rm esc}$ is the escape speed from the stellar surface.  The
terms in the square brackets represent the sum of the change in
potential energy ($1 - R_*/R_{\rm t}$) and the change in kinetic
energy ($0.5 R_*/R_{\rm t}$) of accreting material, minus the work
done on the stellar rotation [$f (R_{\rm t} / R_*)^{1/2}$].  It is
$L_{\rm a}$ that is deposited near stellar surface by accretion, and
thus $L_{\rm a}$ powers energetic accretion phenomena, such as excess
luminosity \citep{konigl91} and a stellar wind.

We suggest that there are a number of possible ways in which some of
this energy will transfer to the open field region of the stellar
corona.  Accretion shocks \citep{konigl91, kastnerea02}, and possibly
magnetic reconnection events \citep{hayashi3ea96}, give rise to X-rays
and UV excesses, which radiate the stellar surface.  Shock heated gas
may diffuse or mix across closed field regions and into the stellar
wind region, and thermal conduction may be significant.
Time-dependent accretion events will excite magnetosonic waves that
may propagate throughout the corona and deposit energy through wave
dissipation.  In general, these processes increase the thermal energy
in the corona, and the details are not necessary for the estimate that
follows.

An MHD wind can be powered by both the rotational kinetic energy of
the star and by coronal thermal energy \citep{washimishibata93}.  We
propose that the thermal component is powered by some fraction
$\epsilon$ of the accretion power, $L_{\rm a}$.  The thermal power in
the wind is approximately $\dot M_{\rm w} v_{\rm s}^2 (\gamma -
1)^{-1}$, where $v_{\rm s}$ is the sound speed near the stellar
surface and $\gamma$ is the polytropic index (i.e., $P \propto
\rho^\gamma$).  Setting this equal to $\epsilon L_{\rm a}$, gives
\begin{eqnarray}
\label{eqn_eng}
\dot M_{\rm w} / \dot M_{\rm a} =
     \epsilon \Gamma_{\rm th}^{-1}
     [1 - 0.5 R_*/R_{\rm t} - f (R_{\rm t}/R_*)^{1/2}]
%
%
\end{eqnarray}
where $\Gamma_{\rm th} \equiv 2(v_{\rm s} / v_{\rm esc})^2 (\gamma -
1)^{-1}$ relates the thermal energy to the gravitational potential
energy.  In reality, the parameter $\Gamma_{\rm th}$ is not
independent of $\epsilon$, since the mechanism(s) by which accretion
energy powers the wind influences the gas temperature (and thus
$v_{\rm s}$), and the location and rate of energy deposition
influences the effective $\gamma$.

This formulation of the problem is advantageous, as it is valid for
wind temperatures ranging from hot, in which thermal pressure
dominates the wind dynamics, to cold, in which magneto-centrifugal
effects dominate (i.e., fast magnetic rotator winds).  The energy
equation (\ref{eqn_eng}) can be combined with the torque equation
(\ref{eqn_feq}) to solve for $f_{\rm eq}$ and $\dot M_{\rm w} / \dot
M_{\rm a}$, simultaneously, for any given coupling efficiency
$\epsilon$ and thermal energy parameter $\Gamma_{\rm th}$.  Assuming
$\gamma = 5/3$, the observed X-ray temperatures and the observations of
\citet{dupreeea05} suggest that the value of $\Gamma_{\rm th}$ for
CTTS's is likely to be in the range 0.3--3.  Adopting the fiducial
values of equation (\ref{eqn_feq}), this likely range of $\Gamma_{\rm
th}$ requires a power coupling efficiency in the range $4\% \la
\epsilon \la 40\%$, to achieve the ratio of stellar mass loss rate to
disk accretion rate of $\dot M_{\rm w}/\dot M_{\rm a} \approx 0.1$ and
an equilibrium spin $f_{\rm eq} \approx 0.09$.  This value of
$\epsilon$ appears reasonable and should help to discriminate between
different possible energy transfer mechanisms.

\section{Synthesis} \label{sec_synthesis}

\begin{figure}
\plotone{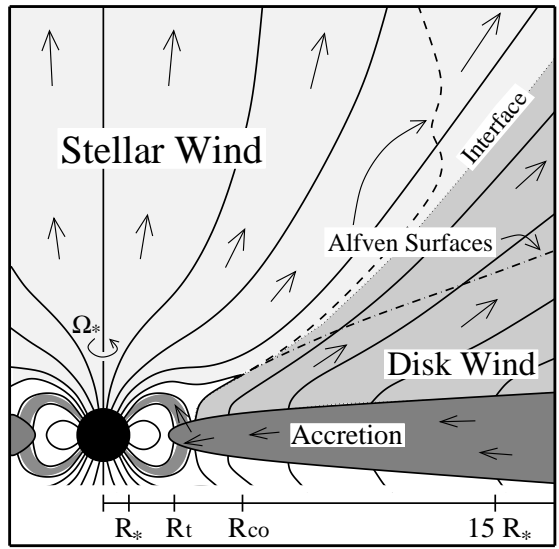}

\caption{Schematic of the star-disk interaction.  The inner edge of
the accretion disk, located at $R_{\rm t}$, is connected to the
stellar magnetic field (solid lines), which regulates the transfer of
matter, energy, and angular momentum to the star (black circle).
Arrows indicate the direction of both matter and angular momentum
flow.  The dashed and dash-dotted lines indicate the location of the
Alfv\'en surfaces in the stellar and disk winds, respectively.
\label{fig_cartoon}}

\end{figure}

Figure \ref{fig_cartoon} illustrates our proposed scenario for the
dynamics and angular momentum evolution of the combined star-disk
system.  This is a synthesis of many results from the literature on
disk winds \citep[e.g.,][]{ouyedpudritz97}, stellar winds
\citep[e.g.,][]{mattbalick04}, funnel flow accretion
\citep[e.g.,][]{romanovaea02}, and the general star-disk interaction.
In the figure, the stellar dipole magnetic field connects only to a
small portion of the disk inner edge, as in ``state 1" of
\citet{mattpudritz05}.  From there, disk material is channeled by the
magnetic ``funnel'' to the polar region of the star, depositing mass,
energy, and angular momentum.  The star is rotating sufficiently
slowly that the corotation radius, $R_{\rm co} \equiv f^{-2/3} R_*$,
is outside the connected region, and the star feels only a spin-up
torque from its interaction with the disk.  At the same time, there is
a powerful wind along the open stellar field.  The stellar wind
Alfv\'en surface (dashed line) is near 15 $R_*$ at mid latitudes, and
crosses the pole at a much larger spherical radius, giving an
effective cylindrical lever arm length, $r_{\rm A}$, of approximately
15 $R_*$.

With an estimate of $r_{\rm A}$, it is possible to consider the
influence of rotation on the wind, since magneto-centrifugal effects
begin to be important when $r_{\rm A}$ is greater than $R_{\rm co}$.
\citet{sakurai85} showed (see his fig.\ 2) that, for $v_{\rm s} /
v_{\rm esc}$ equal to the solar wind value, centrifugal acceleration
is of equal importance with thermal driving when $r_{\rm A} / R_{\rm
co} \sim 100^{1/3}$.  For a star rotating at 10\% of breakup, this
means that equality of thermal and centrifugal effects occurs when
$r_{\rm A} / R_* \sim 22$.  The logical conclusion is that centrifugal
effects will be at least marginally important in CTTS winds when $\dot
M_{\rm w} \sim 10^{-9} M_\odot$ yr$^{-1}$, and may dominate for much
lower values of $\dot M_{\rm w}$ (since $r_{\rm A}$ is then larger) or
faster rotation rates.  Even with marginal centrifugal effects, these
winds should be self-collimated, and most wind parameters (e.g.,
$r_{\rm A}$) depend on $\Omega_*$ \citep{washimishibata93,
mattbalick04}.  Furthermore, at large distances from any magnetic
rotator, wind material possesses angular momentum equivalent to an
amount as if the wind were corotating at $r_{\rm A}$
\citep[e.g.,][]{michel69}. Thus, CTTS winds should rotate at a speed
comparable to that of a disk wind \citep{bacciottiea02, andersonea03}
at observationally resolved distances from the star.

As shown in Figure \ref{fig_cartoon}, a disk wind is present that
extracts angular momentum from the disk.  The Alfv\'en surface of the
disk wind (dash-dotted line) gives an effective lever arm that is a
few times the radius of the footpoint of the field lines from which
the wind flows.  The disk and stellar winds collimate on a scale
larger than the figure.  It is evident that the presence of the
accretion disk is likely to affect the stellar wind.  In particular,
the disk wind can help to collimate the stellar wind
\citep[e.g.,][]{pelletierpudritz92, ouyedpudritz97}, acting as a
hydrodynamic ``channel.''  This would result in shallower magnetic and
thermal pressure gradients and possibly increase $r_{\rm A}$, relative
to the case of an isolated stellar wind.  Finally, the sheared
interface between the two winds is likely to produce observable
signatures from interesting phenomena, such as shocks and
Kelvin-Helmholz instabilities, and there exists a current sheet that
should give rise to magnetic reconnections and particle acceleration.

In essence, the accretion powered stellar wind model solves the
stellar angular momentum problem in the same way that a disk wind aids
angular momentum transport in the disk \citep[][for a
review]{koniglpudritz00}.  Both the star and disk may drive
accretion-powered magnetic outflows that are $\sim 10$\% of $\dot
M_{\rm a}$.  In the case of the disk, the local rotation rate is at
break-up, so a short lever arm is sufficient to provide angular
momentum transport there.  The star, on the other hand, has a much
stronger magnetic field than the disk, resulting in a longer lever
arm, and so an equilibrium spin rate of much less than break-up speed
is possible.

The configuration of Figure \ref{fig_cartoon}, as well as the possibility
that the most significant spin-down torques on the star originate from open
stellar field lines, is well-supported by the numerical MHD
simulations of \citet{goodsonwinglee99} and \citet[][but see
\citealp{romanovaea02}]{vonrekowskibrandenburg04,
vonrekowskibrandenburg05}.  
This picture may also apply to accreting
systems other than CTTSs, such as cataclysmic variables, binary X-ray
pulsars, and accreting black holes.

\section{Conclusion}

We propose that the slow spin of CTTSs is explained by a balance
between the spin-up torque from accretion and the spin-down torque
from a stellar wind (eq.\ [\ref{eqn_feq}]).  In this scenario, some
fraction ($\epsilon$) of the energy released by accretion ultimately
powers a stellar wind with a large mass loss rate ($\dot M_{\rm w}
\sim 0.1 \dot M_{\rm a}$) and rapid angular momentum loss.
Furthermore, we expect that there is a threshold value of $\dot M_{\rm
a}$, below which the contraction of the star is more important than
accretion torques.  At this later time, stellar spin evolution could
be controlled by the interplay between contraction to the main
sequence and a conventional stellar wind.  Thus, an intrinsic spread
in the timescale for the decline of accretion could explain the
distribution of rotational velocities in young clusters
\citep[][]{edwardsea93}, in the same manner as that usually attributed
to disk locking.

Our analysis is free from the problems facing disk-locking models
discussed by \citet{mattpudritz05}.  In particular, the X-wind
\citep[][and subsequent work]{shuea94} and standard star-disk torque
models \citep[][]{ghoshlamb78, konigl91} require a large-scale
magnetic field that is stronger than current observations allow.
These models also assume an unrealistically strong magnetic connection
between star and disk and neglect any torque contribution from a
stellar wind.  In this Letter, we showed that a stellar wind is
capable of providing significant torques, even when the magnetic field
is an order of magnitude weaker than that required by disk-locking
models.  Our estimate of $r_{\rm A}$ (\S \ref{sec_equilib}) suggests
that CTTSs have sufficiently long lever arms, but this calculation
should be made more precise.

Observations of the hot, possibly stellar, outflows can further
constrain our model.  More high-resolution spectroscopy
\citep[e.g.,][]{kastnerea02, dupreeea05} may reveal stellar wind
signatures that will provide better constraints on $\dot M_{\rm w}$,
$\Gamma_{\rm th}$, and $\epsilon$.  Finally, additional measurements
or limits on the large-scale magnetic field
\citep[e.g.,][]{johnskrull3ea99, smirnovea03} would be useful to
constrain the value of $r_{\rm A}$.


\acknowledgements

We are grateful for discussions with Robi Banerjee and Alison Sills,
useful remarks from the anonymous referee, funding from McMaster
University, and a grant from NSERC of Canada.




\end{document}